\input harvmac.tex
\hsize=15.0 true cm
\vsize=22.0 true cm
\hoffset=0 true cm
\voffset=0.0 true cm
\font\bigfc=cmbx10 scaled  1200  

\let \q=`
\def\sqr#1#2{{\vcenter{\vbox{\hrule height.#2pt \hbox{\vrule width.#2pt 
height#1pt \kern#1pt \vrule width.#2pt}\hrule height.#2pt}}}}

\def\gtsim{\mathrel{\hbox{\raise0.2ex
\hbox{$>$}\kern-0.75em\raise-0.9ex\hbox{$\sim$}}}}
\def\ltsim{\mathrel{\hbox{\raise0.2ex
\hbox{$<$}\kern-0.75em\raise-0.9ex\hbox{$\sim$}}}}
\def\AP#1{Ann.\ Phys. {\bf{#1}}}

\def\NP#1{Nucl.\ Phys. {\bf B{#1}}}
\def\PL#1{Phys.\ Lett. {\bf B{#1}}}
\def\PR#1{Phys.\ Rev. {\bf {#1}}}

\def\PREP#1{Phys.\ Rep. {\bf {#1}}}
\def\PRL#1{Phys.\ Rev.\ Lett. {\bf {#1}}}
\def\PTP#1{Prog.\ Theor.\ Phys. {\bf {#1}}}

\def \SPJ{Sov.Phys. JETP}

\def\SPJ#1{Sov.\ Phys.\  JETP {\bf {#1}}}
\def \ds{\displaystyle}

\def \PD {preprint DESY 94-229, HLRZ 94-63}
\def \Romannumeral(#1) {\uppercase\expandafter{\romannumeral#1}}

\def \Romannumeral(#1) {\uppercase\expandafter{\romannumeral#1}}
\def \Rn(#1) {\uppercase\expandafter{\romannumeral#1}}
\def\Fig(#1){$${\overline{\underline{\rm Fig.{\ \ #1}}}}$$} 
\def\Tab(#1){$${\overline{\underline{
 \rm Table\ \  \uppercase\expandafter{\romannumeral#1}}}}$$} 

\def \a{\alpha}
\def \b{\beta}
\def \g{\gamma}
\def \c {\chi}
\def \D{\Delta}
\def \O{\Omega}

\def \ve{\varepsilon}
\def \l {\lambda}
\def \L {\Lambda}

\def \s{\sigma}

\def \th{\theta}

\def \m{ \mu }
\def \n{ \nu }

\def \c {\chi}

\def \l {\lambda}


\def \Ud{U^{\dagger}}
\def \Td{T^{\dagger}}

\def \ft{\tilde f}

\def \fn1{N_{f_1}}
\def \fn2 {N_{f_2}}

\def \Romannumeral(#1) {\uppercase\expandafter{\romannumeral#1}}
\def \Rn(#1) {\uppercase\expandafter{\romannumeral#1}}

\def \det{{\rm det}}
\def \Tr{{\rm Tr}}
\def \ln{{\rm ln}}
\def \ds{\displaystyle} 
\rightline{\hfill YAMAGATA-HET-96-14} \par
\rightline{\hfill KYUSHU-HET-37} \par
\rightline{\hfill SAGA-HE-115}\par
\hfill  Jan., 1997\par
\vskip 1 true cm
\smallskip 
\centerline{\bigfc  Phase Structures of U(2) Gauge Theory}
\centerline{\bigfc   with $\theta$-Term in 2 dimensions }
\vskip 1 true cm
\centerline{Masahiro IMACHI\footnote{*}{ e-mail:imachi@sci.kj.
yamagata-u.ac.jp},
 Takaaki KAKITSUKA\footnote{**}{ e-mail:kaki@aecl.ntt.co.jp}, Norimasa 
TSUZUKI\footnote{***}{ 
 e-mail:tsuz1scp@mbox.nc.kyushu-u.ac.jp}
}
\smallskip
\centerline{and}
\smallskip
\centerline {Hiroshi YONEYAMA\footnote{$^\dagger$}{
 e-mail:yoneyama@cc.saga-u.ac.jp}}
\smallskip
\centerline{$^*$\sl  Department of Physics, Yamagata University, 
Yamagata 990, Japan}
\centerline{$^{**}$\sl NTT Opto-Electronics Laboratories,
 Atsugi, Kanagawa 243-01, Japan}
\centerline{$^{***}$\sl  Department of Physics, Kyushu University, Fukuoka 
812, Japan}
\centerline{\sl $^{\dagger }$ Department of Physics,
            Saga University, Saga 840, Japan }
 ABSTRACT \par
U(2) lattice gauge theory with $\theta$-term in 2 space-time dimensions is
 investigated.
It has non-Abelian real action and Abelian( U(1) type) imaginary action. 
The imaginary action is defined as the  standard $\theta$-term.
 As the effect of
 renormalization group( RG)
 transformation, non-Abelian imaginary action is induced. After many steps of 
RG transformation, non-Abelian
 part will die away. After several steps of RG transformations, renormalized
 action approaches so  called
 heat kernel action. Phase transition is found at $\theta =\pi$ only.\par  
\vskip  4 true cm
\eject
\bigskip
\hbox {\bf $\S 1.$ Introduction }
\smallskip
Much progress has been made in the lattice gauge theory approach in 
these 
twenty years. The systems studied, however, are limited mostly to those 
without 
topological term( =$\theta$-term). This is  due mainly to the difficulty to 
treat
$\theta$-term in numerical calculations in Euclidean space-time.
 Numerical calculations 
are performed with the use of probability weight defined by Euclidean 
action.
 When the system
 is limited to those without $\theta$-term, the probability weight is 
given 
by positive 
quantity. But when $\theta$-term is added to the Euclidean action, we 
obtain complex weight
 due to the nature that $\theta$-term is a pure imaginary quantity in
 Euclidean space-time.
The $\theta$-term, however, is expected to lead to physically interesting
 effects. \par
In 4 space-time dimensions, $\theta$-term leads to strong CP violation 
which 
is severely
 suppressed in real world, although it is not excluded from theoretical
 frame work. The system with $\theta$-term leads to oblique confinement
 or rich phase structures
\ref\Ho { G. $^,$t Hooft, \NP{190}[FS3](1981)455. }\ref\CR{ 
J. L. Cardy and E. Rabinovici, \NP{205[FS5]}(1982)1.} 
\ref\C{J. L. Cardy, \NP{205[FS5]}(1982)17.} 
.\par
In 3 space-time dimensions, it is called Chern-Simons term and is related 
to new physical
 effect, e.g., fractional statistics.
In compact U(1) gauge theory, the system is in the confinement phase, as 
discussed by Polyakov
 due to the monopole excitation. There is a conjecture that the system 
undergoes a phase transition\ref\Fra  {E. Fradkin and F. A. Shaposnik,
 \PRL{66}(1991)267.}. The system is in ``deconfined phase" when
 $\theta$-term is included 
since existence of $\theta$-term
 has the effect to wash out magnetic monopoles and then the system cannot 
be
 in confined phase in
 arbitrary nonzero $\theta$.\par
In two space-time dimensions, the compact U(1) system is in the confinement 
phase. The existence of 
$\theta$-term 
however, leads to deconfinement phase transition
 when  $\theta=\pi$\ref\Col {S. Coleman, \AP{101}(1976)239.}
\ref\HIY{A. S. Hassan,
 M. Imachi and H. Yoneyama,
 \PTP{93}(1995)161.}\ref\HITYU{A. S. Hassan,
 M. Imachi, N.Tsuzuki  and H. Yoneyama,
 \PTP{94}(1995)861.}.\par
The system with $CP^{N-1}$ symmetry in 2 space-time dimensions has much 
similarity to 4 dimensional
 gauge systems( asymptotic freedom, confinement, topological excitations
 etc.) The study of
 $CP^{N-1}$ with $\theta$-term is quite interesting. Especially how
 $\theta$-term affects the phase 
structures will be quite important.
In the  previous paper we numerically studied $CP^1$ model with 
$\theta$-term in two dimensions,
 where we obtained $\theta$- dependence of the partition function 
through the measurements of
 topological charge( $Q$) distribution$P(Q)$\ref\HITY{A. S. Hassan,
 M. Imachi, N.Tsuzuki  and H. Yoneyama,
 \PTP{95}(1996)175.}.
The fourier transform of this distribution gives us the partition 
function $Z(\theta)$\ref{\Wi}{U. -J. Wiese, \NP{318}(1989)153.}.
 We saw  first order phase transition  in strong real coupling regions and 
weaker transition in
 weak coupling regions (but we could not conclude that it belongs to 2nd order 
phase transition).\par

Schierholz has pointed out that $CP^{N-1} (N=4)$ model in 2 dimensions 
undergoes  
first order transition
 at some values of $\theta$, depending on the real coupling . In strong 
(real)
 coupling, $\theta_c$ 
is given by $\pi$ and $\theta_c$ deviates from $\pi$ at those $\beta$ 
greater 
than some value. 
He conjectured that $\theta_c$ goes to zero at weak coupling limit
 ( $a \rightarrow 0$). According to him,
 the continuum limit value of $\theta$ would be  zero and this would be  
the dynamical 
explanation for the smallness 
of physical $\theta$\ref \SC{G. Schierholz,`` $\th$ Vacua, 
Confinement and The Continuum Limit", \PD.}\ref \SC {G. Schierholz,
 hep-lat/9409019.}.\par
The study of $CP^{N-1}$ system or gauge systems with $ \th$-term will be
 quite important to
 understand  really the reason of vanishingly small value of $ \th$ in four 
dimensional QCD. There will
 be various approach to study 
the system with $\theta$-term. One of them is numerical simulation, where 
the 
difficulty of complex
 Euclidean action and thus complex Boltzmann weight is overcome by first 
obtaining topological charge($Q$)
 distribution $P(Q)$ and then by calculating $Z(\theta)=\Sigma_Q P(Q)
 e^{i \theta Q/2\pi}$\ref{\Wi}{U. -J. Wiese, \NP{318}(1989)153.}.
 Another method to study the phase structure  
is to apply real space renormalization group method to the 
system\ref{\MI}{A. A. Migdal, \SPJ{42}(1976)413, 743.}
\ref{\KA}{ L. P. Kadanoff, \AP{100}(1976)359.}
\ref {\BT}{K. M. Bitar, S. Gottlieb and C. K. Zachos, \PR 
{26}(1982)2853, \PL{121}(1983)163.}
\ref\IKY{M. Imachi, S. Kawabe and H. Yoneyama, \PTP{69}(1983)221, 1005.}. 
In this approach, the information of
 the phase structure is obtained through the analysis of renormalization 
group flow of various 
couplings($\beta,\  \theta$ etc.)\ref\WK { K. G. Wilson and J. B. Kogut,
 \PREP{12C}(1974) 75.}.\par
In this paper we apply real space renormalization group method to 2 
dimensional U(2) lattice 
gauge theory\ref \KTS { T. G. Kovacs, E. T. Tomboulis and Z. Schram,
\NP{454}(1995)45. }.
 The reason to choose U(2) symmetry is that we are interested 
in the system with both
 non Abelian( SU(N)) and Abelian( U(1)) part. In order to have nontrivial 
$\theta$-term 
we have to study 
the gauge group with Abelian part, because standard imaginary action 
Tr$\ve_{\m \n} F_{\m \n}$
becomes nontrivial when gauge group has U(1) part.
 The symmetry SU(N), without U(1) part,  on the other hand leads 
to trivial $\theta$-term, since Tr$\ve_{\m \n} F_{\m \n}$
 is zero. So we have chosen U(N) symmetry. The group U(2) is the simplest
  among U(N)($N \ge 2$)'s. 
 In U(2) gauge symmetry, we will pay attention to the interplay 
between Abelian  and 
non-Abelian SU(2) effects. To see nontrivial effect, we choose the real 
coupling term as 
the fundamental representation $\beta_{l_1 l_2}=\beta _{11}( l_1$ and 
$l_2$ 
are U(1) and SU(2)
 part of U(2) symmetry, respectively and $l_1=1$ and $l_2=1$ means that 
the  action belongs 
to fundamental representation of  both U(1) and SU(2) part).\par

In the choice of $\theta$-term action, there are two possibilities 
according
 to possible U(N) gauge invariants; det$U$ and Tr$U$, where $U$ 
is an element of U(N) group  belonging to 
fundamental representation. One is ``standard $\theta$-action" which is
 defined as
 ${1 \over 2}{\theta \over {2\pi}}(\ln \ \det U-{\rm c.c.})$ 
corresponding to the continuum limit form $i {\theta \over {2\pi}}F_{\mu 
\nu}
\varepsilon_{\mu \nu}$.
The other choice  is ``Wilson $\theta$-action" defined as ${1\over 2}{\theta 
\over 
{2\pi}}(\Tr U- {\rm c.c.})$
 corresponding to imaginary version of ``Wilson action"( real Wilson
 action is defined
 as ${1\over 2}\beta (\Tr U+{\rm c.c.})$).\par
In this way, bare actions we consider in this paper  are given by\par
\item{(i)}real action which has both U(1) and SU(2) part of U(2) group\par
\item{(ii)}imaginary action which has only U(1) part.\par
 
We specify the irreducible representation(IR) of U(2) by two integers 
$l_1$ and $l_2$ corresponding 
to U(1) and SU(2) part, respectively.
 Trivial representation is given by $l_1=0$ and $ l_2=0$. The set of real 
and imaginary 
couplings is denoted as $\beta_{l_1 l_2}$ and $\gamma_{l_1 l_2}$ 
respectively. Bare action is
 defined by 
\item{(1)}  $ \beta_{11}\not= 0$ and $ \beta_{l_1 l_2}=0$ for other 
representations 
and
\item{(2)}  bare imaginary 
action contains only U(1)
 part( (i) $\alpha x$ (standard $ \theta$-action ) or (ii) $\alpha \sin 
x$ 
(Wilson $\theta $-action)
 respectively), where $x$ is defined as the U(1) variable of U(2) group
(${\ln \ \det} U-{\ln\  \det} \Ud= 2 i x$ and 
${\Tr }U -{\Tr} \Ud =2 i \sin x$). In these actions,
$ \gamma_{l_1 l_2}$ with $l_2\not=0$ are zero. \par
Although bare imaginary action is given as the function of U(1) variable 
$x$ only, 
the RG 
transformation induces nontrivial SU(2) part, and we have, for 
example,  $\gamma_{11}\not= 0$
 after RG transformations. On the other hand non zero $\beta_{20}$ ( 
originally zero) is induced
 as the RG effect. RG flow of ($\beta_{20},\b_{40}$) shows interesting 
behavior.\par
In the case of standard  $\th$-term action, we find phase transition at
 $\th =\pi$.
 The RG flow projected onto ($\b_{20}$ and $\b_{40}$) plane is
 controlled by a fixed point specified by
$\th=\pi$.  For $\th=$ very close to $\pi$ but $\th\not=\pi$, the RG flow 
is 
strongly affected by the $\th=\pi$
 fixed point, but after many steps of RG transformation, the RG flow 
finally 
goes to strong coupling fixed point
 ($\b_{20},\b_{40})=(0,0)$.\par
Deconfinement transition at $\th=\pi$ is limited 
to the coupling belonging  trivial SU(2) representation and 
fundamental representation of U(1), i.e.,
 $ \b_{20}$. The coupling belonging to non trivial SU(2) representation, 
e.g.,
 $\b_{11}$ does not
 show deconfinement transition even at $\th=\pi$.\par
\vskip 1cm
\eject
\hbox {\bf $\S 2.$ Formulation}
\smallskip
\hbox {\bf $\S 2.1$ irreducible characters }
\smallskip
In this paper we consider a gauge group, in 2 space-time dimensions, which
 has both non-Abelian part and  
 $\th$-term(  U(1) part). As
 the simplest of such group, U(2) group is adopted and  interplay between
 Abelian and 
non-Abelian part will be investigated. \par
We denote the element of fundamental representation of U(2) group as $U$ 
and
$$ U=T \L \Td        \eqno(2.1)$$
where $T$ is a unitary matrix diagonalizing $U$ to $\L$ like
$$ \L=
  \pmatrix{ e^{i \phi_1}  &      0       \cr
               0            & e^{i \phi_2}  } 
 =e^{i \a_1/2} 
\pmatrix{
           e^{i \a_2/2}  &   0\cr
              0     &        e^{-i \a_2/2}   
}.\eqno(2.2)$$
Two parameters $\phi_1, \phi_2 (-\pi\leq \phi_1, \phi_2 \leq \pi)$ of 
the diagonal part are replaced by two parameters $\a_1, \a_2$ as
$$ \eqalign{
    \a_1=& \phi_1+\phi_2 \cr
    \a_2=& \phi_1-\phi_2.\cr
}\eqno(2.3) $$
Characters $ \Tr U$ of general irreducible  representations of U(2) group 
are given by 
$$ \chi_{\l_1 \l_2}(\phi_1, \phi_2)=\D_{\l_1 \l_2}(\phi_1, \phi_2)/\D_{0 
0}(
\phi_1, \phi_2)  \eqno(2.4)$$
where determinant is defined as 
$$\D_{\l_1 \l_2}(\phi_1, \phi_2)\equiv
 \det 
\pmatrix{ e^{i \phi_1 (\l_1+1)}  &  e^{i \phi_1 \l_2 }   \cr
         e^{i \phi_2 (\l_1+1)}  &  e^{i \phi_2 \l_2}   }   \eqno(2.5)$$
and 
$$ \D= \D_{0 0}=
\det
\pmatrix{
        e^{i \phi_1}   &   1   \cr
        e^{i \phi_2}   &   1   }.    \eqno(2.6)$$
Integers $\l_1$  and $\l_2$ denote the length of the first  and the 
second row
of Young tableau(see ref. Drouffe et al.\ref\DZ { J-M. Drouffe and J-B.Zuber, 
\PREP{102}(1983) 1.}).
 Note that $\l_1$ and $\l_2$ are 
allowed
to take both positive and negative values. Only constraint is $\l_1 \geq  
\l_2$.
 Dimension of IR 
is 
$$d_{\l_1 \l_2}=\l_1 -\l_2+1      \eqno(2.7)$$.\par
We will introduce another notation referring to U(1) and SU(2) part; 
$$\cases
{  \ds{l_1=\l_1+\l_2  }\cr
   \ds{ l_2=\l_1-\l_2 }  
}    \eqno(2.8)$$
$$ \cases
{  \ds{x=\a_1= \phi_1+\phi_2}\cr
   \ds{           y=\a_2= \phi_1-\phi_2}}\eqno(2.9)$$
  where $l_1(l_2)$ and $x(y) $ denotes the U(1)( SU(2)) part.
As is evident from this definition, $l_1$ and $l_2$ are not completely
 independent but they obey  constraints
$ l_1+l_2=$ even integer and $l_2 \geq 0$.\par
Explicit form of the character is
$$\chi_{l_1 l_2}(x, y)=e^{i {l_1 \over 2}x} 
             {\sin(l_2+1){y \over 2} \over \sin{y\over 2}}      \eqno(2.10)$$
$$ d_{l_1 l_2}=\chi_{l_1 l_2}(0,0)=l_2+1.                       \eqno(2.11)$$
Two integers $l_1$ and $l_2$ denote   $2 q$( $q$ is 
``U(1) charge") 
 and  $2I$($I$ denotes the magnitude of ``isospin"), respectively.\par
 Invariant measure is given by\ref\Weyl {H. Weyl, ``The Classical Group"( 
Princeton Univ. Press( 1939).} 
$$d \O=dT{1\over{8\pi^2}}d\phi_1 d\phi_2 \D \D^*,              \eqno(2.12) $$
 where $\D^*$ is the complex conjugate of $\D$.\par
Fundamental representation is defined as a single box in the Young tableau,
 i.e., $\ (\l_1, \l_2)=(1,0)$, thus $(l_1, l_2)=(1,1)$. \par
\vskip 1cm
\hbox {\bf $\S 2.2$ Migdal Renormalization group}
\smallskip
Partition function $Z$ is defined as an integral
$$     Z=\int d\Omega F(L, u)$$
at scale $L$. The integrand $F$ is given by 
$$                 
F(L,u)=\exp(s(u) + i v(u)),                                       
 \eqno(2.13)$$
 where $u$ denotes an element of U(2) group. This integrand $F$ will be
 expanded by irreducible characters as,
  $$ F(L, u)=\sum_{l_1 l_2} d_{l_1 l_2} \c_{l_1 l_2}(u) \ft_{l_1 l_2}(L).   
                                                               \eqno(2.14)$$
 Real and imaginary part
 of action $s(u)$ and $v(u)$, respectively, are also expanded  into a series as
$$\cases
{
  \ds{s(L, u)=\sum_{l_1 l_2} \b_{l_1 l_2} ({\c_{l_1 l_2}(u) \over d_{l_1 
l_2}}-1)} \cr
  \ds{v(L, u)=\sum_{l_1 l_2} \g_{l_1 l_2}\c_{l_1 l_2}(u)},
}
                                                             \eqno(2.15)$$
where $d_{l_1 l_2}=l_2+1$  is the dimension of IR( irreducible representation) 
$(l_1, l_2)$.\par
Using the orthonormality of the characters, RG transformation is 
defined\MI.
Change of scale $L$ to $\l L$ leads to
$$ F(\l L,u)=\sum_{l_1 l_2} d_{l_1 l_2} \c_{l_1 l_2}(u)
 (  \ft_{l_1 l_2}(L))^{\l^2}.\ \ \                          \eqno(2.16)$$ 
After $t$ steps of RG transformation,
$$F(\l^t L,u)=\sum_{l_1 l_2} d_{l_1 l_2} \c_{l_1 l_2}(u) (\ft_{l_1 
l_2}(l))^{\l^{2t}}.                                       \eqno(2.17) $$
The long distance behavior is determined by this equation at $t\rightarrow 
$large.\par
The change of coupling constants with scale transformation is determined 
by the character expansion of
 $F(\l L, u)$ as
$$F(\l L, u)=\exp\{ \sum_{l_1 l_2}\b_{l_1 l_2}(\l L) ({\c_{l_1 l_2}(u) 
\over d_{l_1 l_2}}-1) +i\sum_{l_1 l_2}
\g_{l_1 l_2}(\l L) \c_{l_1 l_2}(u)\}\ \                        \eqno(2.18)$$
where left hand side is given by eq.(2.16). Equations (2.16) and
 (2.18) define 
the change 
$$ \{\b_{l_1 l_2}(L), \g_{l_1 l_2}(L)\} \rightarrow \{ \b_{l_1 l_2}(\l L), 
\g_{l_1 l_2}(\l L) \}                                         \eqno(2.19)$$
 and successive transformations give us the renormalization flow of 
coupling
 constants.\par
The real part of bare action is chosen as Wilson action with $\b_{11}^b 
\not= 0, \b_{l_1 l_2}^b=0$ for all
 of other representations. For the imaginary part of the bare action, we can  
consider two possibilities;(1)
``standard" imaginary action defined by 
$$  v_{\rm st}(x)=i \a (\ln\  \det U-\ln\  \det \Ud)/2i=i \a x    \eqno(2.20)$$
 and (2) ``Wilson"  imaginary 
action defined by 
$$    v_{\rm Wil}(x)=i \a\  (\Tr\  U- \Tr\  \Ud)/2i 
\ \ \ =i \a \sin x, (\a \equiv \th/2\pi).                    \eqno(2.21)$$
 Both of these approach $i \a x$ in the 
naive continuum limit.\par
In this paper standard imaginary action will be investigated. Due to the 
character expansion,
$$\eqalign{
 ``x"=i\{& -\c_{20}(x)+{1\over 2}\c_{40}(x)-{1\over 3}\c_{60}(x)+\cdots \cr
     &  +\c_{-20}(x)-{1\over 2}\c_{-40}(x)+{1\over 
3}\c_{-60}(x)-\cdots\}\cr
   =2\{&\sin x-{1\over 2}\sin 2 x+{1\over 3}\sin 3 x-\cdots\}.
}                                                                \eqno(2.22)$$
Namely,  the character expansion coefficient $\g_{l 0}$ of $``x"$ is
$$\cases
{
\ds{ \g_{l 0}={2i\over l}(-1)^{l/2}} (l\not= 0)\cr
\ds{ \g_{00}=0}
}                                                      \eqno      {\rm  (I)}$$
where
$$ ``x"=
\cases
{
 \ds{ x-2 \pi }     &{\rm for}  $ \pi<  x \leq 2\pi$ \cr
 \ds{ x }             & {\rm for} $- \pi <x \leq \pi $\cr
 \ds{x+2\pi }        & {\rm for} $- 2\pi \leq x \leq -\pi$.
}
                                                     \eqno(2.23)$$
In the character expansion, we should be careful with the treatment of 
$x=\phi_1+\phi_2$ and $y=\phi_1-\phi_2$, 
because originally, $\phi_1$ and $\phi_2$   range over from $-\pi$ to $\pi$.
We present an example of integration over Haar measure.
$$\ft_{l_1 l_2}= \int d\O \c_{l_1 l_2}^*(x,y)F(x,y)=\ft^{(1)}_{l_1 
l_2}+\ft^{(2)}_{l_1 l_2}                                         \eqno(2.24)$$
where 
$$\ft^{(1)}_{l_1 l_2}= \int _\pi^{2\pi} dy (1-\cos y)\int 
^{2\pi-y}_{-2\pi+y} dx F(x,y)\c^*_{l_1 l_2}                     \eqno(2.25)$$
$$\eqalign{
\ft^{(2)}_{l_1 l_2}=& \int _{0}^{\pi} dy (1-\cos y) \{ \int 
_{\pi}^{2\pi-y}F(x-2\pi,y)
                              +\int^\pi_{-\pi}F(x,y)\cr 
                   &+\int ^{-\pi}_{-2\pi+y} F(x+2\pi,y)\}\c^*_{l_1l_2}(x,y)dx 
.\cr
}                                                                 \eqno(2.26)$$
Two typical examples are presented below.\par
\vskip 1cm

\item{\bf Example 1}.\par
When $F(x,y)=e^{i \a x}$, i.e., pure imaginary action, $l_1$ is limited to 
even integer since $l_2$ is zero.
$$\ft_{l_1 0}= \int d \O F(x,y)\c ^*_{l_1 0}(x,y)               \eqno(2.27)$$
Then

$$\cases
{\ds{ \ft ^{(1)}_{l_1 0}}
       =&$ {1\over 8 \pi^2}\int _ \pi^{2\pi} dy(1-\cos y)
      \int_{-2\pi+y}^{2\pi-y} dx e^{i\a x-i{l_1\over 2}x}  $ \cr
\ds{ \ft ^{(2)}_{l_1 0}}
      =&${1\over 8 \pi^2} \int _0^{\pi} dy (1-\cos y)
\{ \int^{2\pi-y}_\pi dx e^{i\a (x-2\pi)} e^{-i l_1 x/2 }  $\cr
\ds{} &$ + \int ^\pi _{-\pi} dx e^{i\a x}e^{-i l_1 x/2}+ \int 
_{-2\pi+y}^{-\pi} dx 
e^{i \a (x+2\pi)} e^{-i l_1 x/2}\}$}                     \eqno(2.28)$$
which gives
$$\cases
{\ds{ \ft ^{(1)}_{l_1 0}}=&${1\over 8 \pi^2}\int _0^{\pi} dy (1-\cos y)
\int^y_{-y} dx e^{i (\a-{l_1\over 2})x} $ \cr
\ds{ \ft ^{(2)}_{l_1 0}}
      =&${1\over 8 \pi^2}\int _0^{\pi} dy (1-\cos y)
\{2\int ^\pi_{-\pi} dx e^{i(\a- {l_1\over 2})x}-\int^y_{-y} dx 
e^{i(\a-{l_1 \over 2})x}\}$.
}                                                        \eqno(2.29)$$
And finally we obtain
$$\eqalign{
\ft_{l_1 0}
=&\ft^{(1)}_{l_1 0}+\ft^{(2)}_{l_1 0}\cr
=&{1\over{ 4\pi^2}} \int_0^{\pi} dy 
(1-\cos y)\int ^\pi_{-\pi} dx e^{i(\a-{l_1\over 2})x}\cr
=&{1\over {2\pi}} {1 \over{\a-{l_1\over 2} } }\sin((\a-{l_1\over 2})\pi)\cr
=& A_{l_1} (\a).\cr
}\ \                                              \eqno          {\rm (II)}$$
This is just the same form as the one  obtained in U(1) gauge theory with pure 
$\th$-term.\par
\par

\eject
\item{\bf Example 2}.\par
For 
$$F(x,y)=``x"=
\cases
{
 \ds{ x-2 \pi }     &{\rm for}  $ \pi\leq  x \leq 2\pi$ \cr
 \ds{ x }             & {\rm for} $- \pi \leq x \leq \pi $\cr
 \ds{x+2\pi }        & {\rm for} $- 2\pi \leq x \leq -\pi$,
}
                                                                \eqno(2.30)$$
$$\ft_{l 0} = {2 \over i}{d A_l (\a)\over d \a} \vert_{\a=0}    \eqno(2.31)$$
where $A_l (\a)$ is defined in the above example 1. We have
$$\ft_{l 0}=\cases
{
\ds{0 }  &for $l=0$ \cr
\ds{{2 i\over l}(-1)^{l \over 2}} & for$l\not= 0,l=$even.
}\ \ \                                                    \eqno  {\rm (III)}$$
This coincides with the result (I).\par
\vskip 1cm
\vskip 1cm
\vskip 1cm

\hbox {\bf $\S 3.$ Results}
\smallskip
We adopt fundamental representation as the bare real action, i.e., Wilson 
action and ``standard $\th$-term action" as  the bare imaginary action.
$$F=\exp\{\b(\cos \phi_1 +\cos \phi_2-2)+i\a (\phi_1+\phi_2)\},    \eqno(3.1)$$
where $\b$ denotes $\b_{11}$ since
$$\c_{11}=e^{i{x\over 2}} 2 \cos{y\over 2}                      \eqno(3.2)$$
$$\c_{11}+\c^*_{11}=4\cos{x\over 2}\cos{y\over 2}=2(\cos \phi_1 
+\cos{\phi_2}).                                               \eqno(3.3)$$
This bare action contains both U(1) and SU(2) part in the real action and 
only U(1) part in the imaginary action.
Although SU(2) part is not contained in the imaginary action at bare 
level, renormalization transformation leads 
to SU(2) part also in the imaginary action as a consequence of 
interference between real and  imaginary action.\par
\eject
\hbox {\bf $\S 3.1$ $\a=0$( $\th=0$) case(pure real action)}
\smallskip

When $\a=0( \th=0)$ we have simple form of character expansion 
coefficients,
$$F=\sum_{l_1+ l_2={\rm even}} \ft_{l_1 l_2}\c_{l_1 l_2}(x,y)d_{l_1 l_2}
                                                         \eqno(3.4)$$
$$\ft_{l_1 l_2}={e^{-2 \b}\over d_{l_1 l_2}}\det_{ij}(I_{\l_i-i+j} 
(\b)),                 \ \ \ \ \ \                       \eqno(3.5)$$
where 
$$\cases{
\ds{\l_1}=&$(l_1+l_2)/2$ \cr
\ds{\l_2}=&$(l_1-l_2)/2$.
}                                                       \eqno(3.6)$$
and  $I_n (\b)$denotes the Modified Bessel functions.\par
Bare action contains $\b^b_{11}\not=0$, others=0. As the result of 
renormalization transformation
 all coefficients $\b^r_{l_1 l_2}$ (renormalized) appear. We pick up 
$(\b_{11}, \b_{22})$ from 
these infinite
 dimensional coupling constant space. The renormalization flow starting 
from various bare 
couplings converges 
to a trajectory in $(\b_{11}, \b_{22})$ plane(Fig. 1 and Fig. 2).
\Fig( 1 ) \Fig ( 2)
They approach  so called heat 
kernel\ref\MO {  P. M. Menotti and E. Onofri, \NP{190}[FS3](1981)288. }. In the weak
 coupling limit
 ($\b \rightarrow \infty$), modified Bessel function is given by
$$I_n(\b) \sim {e^\b \over{\sqrt{ 2 \pi \b}}}e^{-{n^2\over{2 \b}}} \eqno(3.7)$$
and we have in the case of U(2) gauge group, with the use of (3.5),
$$ F^{h.k.}_{\rm U(2)}=\sum_{l_1+ l_2={\rm even}} d_{l_1 l_2} \c_{l_1 l_2} 
\ft^{h.k.}_{l_1 l_2}
=\sum d_{l_1 l_2} \c_{l_1 l_2}
             e^{-{1\over{2\b}}({1\over 2}l_1^2+l_2({l_2\over 2}+1))}/2\pi 
\b^2                                                           \eqno(3.8)$$
Namely,
$$\eqalign{
F^{h.k.}_{\rm U(2)}
 =  &\sum_{l_1, l_2 {\rm all}} {1\over 2}(1+(-1)^{l_1+l_2})d_{l_1 
l_2}\c_{l_1 l_2}(x,y) \ft ^{h.k.}_{l_1 l_2}\cr
 = &\sum_{l_1, l_2 {\rm all}} \sum^1_{\nu=0} d_{l_1 l_2}{1\over 2} e^{il_1 
x_{\nu}/2}{\sin(l_2+1){y_{\nu}\over 2}
        \over \sin{y_{\nu}\over 2}} \times e^{-{1\over{2\b}}C_{l_1 l_2}}\cr
}                                                               \eqno(3.9)$$
where 
$$\cases{
\ds{x_{\nu}}=&$x+2\pi \nu$ ,\cr
\ds{y_{\nu}}=&$y-2\pi \nu$ ,
}                                                            \eqno(3.10)$$
with $\nu=0, 1$.\par

In this way the character expansion coefficient is written as
$$ \ft^{h.k.}_{l_1 l_2}={1\over{2\pi \b^2}}e^{-{1\over {2 \b}}C_{l_1 
l_2}}                                                         \eqno(3.11)$$
where 
$$C_{l_1 l_2}={\rm quadratic\   Casimir\  operator\  of\ 
 U(2)}={1\over 2}l^2_1+l_2({l_2\over 2}+1).                      \eqno(3.12)$$
In strong coupling region, heat kernel (3.9) gives the relation between 
coupling constants. For example we have
$$\b_{22}\sim -{3\over 8} \b_{11}^2                            \eqno(3.13)$$
at $\b <<1$. In Fig. 1 and Fig. 2, coupling constant relation obtained 
by heat kernel (3.9) is shown. The actual trajectory in strong coupling 
region is quite close to (3.13).\par
After $t$ steps of renormalization transformation
$$F^{h.k.}(\l^t a,x,y)=\sum_{l_1+ l_2={\rm even}} d_{l_1 l_2} \c_{l_1 
l_2}(x,y)(\ft^{h.k.}_{l_1 l_2})^{\l^{2 t}}                      \eqno(3.14)$$
$$( \ft^{h.k.}_{l_1 l_2})^{\l^{2 t}}\propto e ^{-{\l^{2t}\over{2\b(a)}} C_{l_1 l_2}}\equiv 
e^{- {1\over{2\b(\l^t a)}} C_{l_1 l_2}}                      \eqno(3.15)$$
Namely,  coupling constant $\b$ in heat kernel  is transformed, after $t$ 
steps of renormalization transformation,
  to simply $\b/\l^{2t}$, (at each transformation $\beta$ is divided 
by $\l^2$), i.e., the functional form is not 
changed but only the value of 
$\b$ is changed.\par
\vskip 1.0cm
\hbox {\bf $\S 3.2$ $\a \not= 0$( $\th\not= 0$) case(imaginary action)}
We introduce standard $\theta$-term action as the bare action. In this
 case bare 
imaginary
 coupling constants are 
$$\g^b_{l_1 0}={i(-1)^{l_1/2 }\over l_1} 2 \a ,\ \  \g^b_{0 0}=0, \g^b_
 {l_1 l_2}=0\  {\rm for}\  l_2\not= 0.                         \eqno(3.16)$$
Then renormalized couplings are non zero, $\g^r_{l_1 l_2}\not= 0$ even for 
$l_2\not= 0$, e.g., $\g_{11}, \g_{22} $ etc. are induced by 
renormalization 
effects through interference between $\b_{11}$ and $\g_{l_1 0}$.\par
\vskip 1cm
In the following, some characteristics of the RG flow will be  discussed  
 in detail.\par

\smallskip
\item {1)}$(\b_{11}, \b_{22})$ flow; The flow projected onto $(\b_{11},
 \b_{22})$ plane depends on bare $\th$, but flows starting from different 
$\b$ 
converge to a unique trajectory  independently of bare parameters. As 
$\th$ approaches
 $\pi$, trajectory is much affected by the fixed point $\th=\pi$. The 
trajectory, however, does not stay at finite $(\b_{11}, \b_{22})$, but 
finally approaches 
 the strong coupling limit (0, 0) at large $t$(= RG 
transformation step) . It shows that the charge of fundamental
 representation of
 U(2) group is in the 
confinement phase even at $\th= \pi$. Main reason is that SU(2) part of 
U(2) plays important role for the charge of fundamental representation.\par
\Fig( 3)
\item{2)}$(\b_{20}, \b_{40})$ flow; Originally $\b^b_{l_1 0}=0$ for all 
$l_1$ in our
 set of bare parameters. These($\b^r_{l_1 0}$'s)  are induced by RG 
transformation. In the 
first $t$ steps
$(<t_0)$, which we call region I, renormalized couplings $(\b^r_{20}, 
\b^r_{40})$ are 
strongly affected by the fixed point $ (\th=\pi)$. As we change bare 
$\th$, the influence of 
the fixed point becomes stronger as $\th $ approaches $\pi$. After $ t_0$ 
steps, flows start
 to move to infrared 
 fixed point $(\b^r_{20}, \b^r_{40}) =(0, 0)$ for $\th^b\not= \pi$.
At $\th=\pi$, $(\b^r_{20}, \b^r_{40})$ approaches fixed point (1/2,$ 
-1/4$ ) as
 $t\rightarrow $ large.
Namely, at $ \th=\pi$ , only region I exists but region II does not.\par
\Fig( 4)

We have from eq.(II) in section 2,
$$\ft_{00}(\a)=\ft_{20}(\a)                               \eqno(3.17)$$
when $\a={1\over 2}(\th=\pi)$. The other coefficients are smaller than these 
two.
$$\ft_{00}({1\over 2})=\ft_{20}({1\over 2})>\ft_{-20}({1\over 
2})=\ft_{40}({1\over 2})>\cdots                               \eqno(3.18)$$
and RG transformation leads to the result that large distance behavior is 
dominated by $\ft_{00}=\ft_{20}$ for $\a=1/2$.
$$F\sim \ft^{\l^{2t}}_{00}(1+e^{i x})\propto 2\cos {x\over 2} e^{i{x\over 
2}}                                                     \eqno(3.19)$$
at $t\rightarrow \infty$   and  $\l>1$. This gives
$$\eqalign{
s=& \ln (2 \cos {x\over 2})\cr
v=&{x\over 2}.\cr
}                                                   \eqno(3.20)$$
Renormalized $\th$-term action is  exactly  given by ${1\over 2}x$( fixed point 
action).
Namely, for $\a^b=1/2$, RG flow in $(\b_{20},\b_{40})$ plane converges to
$ ({1 \over 2},{-1\over 4})$ point,
because fixed point action $s=\ln (2 \cos {x\over 2})$ leads to 
$\b_{l_1 0}={2\over |l_1|} (-1)^{{l_1\over 2}+1} (l_1 \not= 0)$.\par

Charge in (2, 0) 
representation is deconfined at $\th=\pi$ by $\th$- term. (In pure U(1) 
theory we showed 
fundamental representation charge is deconfined.)  $\th$-term works as 
the back ground field to 
 the U(1) part of U(2) group and background field just cancels the 
electric field produced by the 
 U(1) charge of Wilson loop.\par
\item{3)}$({\rm Im }\g_{11}, \b_{11})$ flow; Originally $\g_{11}=0$, i.e., 
the coupling with nontrivial SU(2) part was absent in the bare action.
 The coupling $\g_{11}$ is induced 
as RG effect for 
nonzero $\th$'s($0<\th<\pi)$, while $\g_{11}$ is zero for $\th=0$.
 Only small 
$|\g^r_{11}|$ is induced
for $\th \rightarrow \pi$. For $\th^b=\pi$, $\g^r_{11}=0$, because at 
$\th=\pi$, $\th$-term
 remains fixed point. In that case $\g^r_{l_1 0}\equiv {2 i\over 
l_1}(-1)^{l_1/2}$, and,
 $\g^r_{l_1l_2}=0$  for all $l_2\not= 0$. In Fig. 5.  $({\rm Im }\g_{11}, \b_{11})$
 flow 
is shown in Fig. 5.\par
\Fig( 5)
\item{4)}string tension; The string tension is defined by
$$\s_{l_1 ,l_2} =-{1\over L^2}\ln{{\int {d\O} \chi_{l_1, l_2}^*(u)F(L, u)}
\over{\int 
{d\O}F(L, u)}}\eqno(3.21) $$

   The string tension with nontrivial SU(2) part 
, i.e., string tension $\s_{11}$ coming from the  Wilson loop of 
IR(irreducible representation) $(l_1, l_2)=(1, 1)$,
decreases as $\theta$ approaches $\pi$ and $\s_{11}\rightarrow $ finite 
 as $\th \rightarrow \pi$. String 
tension for 
 representation  (2, 0) Wilson loop, i.e., string tension with trivial
 SU(2) part,
 $\s_{2 ,0}\rightarrow 0 $ as 
$\th\rightarrow \pi$.
This again shows that  (2, 0) charge is deconfined at 
 $\th=\pi$ but  (1, 1) is
not  even at $\th=\pi$. In the latter
 case, SU(2) part 
drives the system always to the confinement phase.\par
 Free energy is also shown. Free energy is
$$      -\{\ln \int d\O F(L, u) \}/L^2     \eqno(3.22)$$
 It is nearly proportional to $\th^2$.\par
\Fig( {6(a), (b)})

\item{5)}$({\rm Im} \g_{20}, {\rm Im}  \g_{40})$ flow; Qualitative difference is seen between 
$t<t_0$( region I), and $t> t_0$(  region II).
It is attracted by the $\th=\pi$ fixed point  and moves very 
slowly due to this attraction for $t<t_0$.
 Note that the flow for this pair of couplings starts from the nearest
 point to  $\th=\pi$ fixed point. It   moves on a 
parabola to $(\g_{20}, \g_{40})=(0, 0)$ fixed point for $t> t_0$. 
 Flows for various 
$\th$ are plotted in Fig. 7.
 At $\th=0.999\pi$ the flow is strongly attracted to the 
$\th=\pi$ fixed point and stays rather long time near  this point( 
region I), but finally trajectory
 moves toward (0, 0)  fixed point( region II).
 For $\th=\pi$ bare theory, RG  flow does 
not 
move  at all( the system is always in region I) and  $({\rm Im} \g_{20}, 
{\rm Im} \g_{40})$  stays at  $(-{1\over 2}, {1\over 4})$
\par
\Fig( 7)

\item{6)}$({\rm Im}  \g_{11}, {\rm Im} \g_{22})$ flow; Originally the value of $(\g_{11},
 \g_{22})$ is (0, 0). 
But they( $\g_{11}$ and $\g_{22}$) are induced by RG transformation 
by the interference between real 
$\b_{11}$ Wilson term
 and imaginary $\g_{l_1 0}$'s. As seen in Fig. 8, the flow starts from
 (0, 0) point
and
$|{\rm Im}  \g_{11}|$ and $|{\rm Im}   \g_{22}|$ becomes lager, in region I and then get
 back to the strong 
coupling
fixed point( region II).\par
\Fig( 8)
\item{7)}Change of $\th$-term under RG; In this paper we adopted 
``standard "
 action for imaginary($\th$)-term.As in Fig. 9, originally the imaginary action is 
given by linear dependence on $x$ with $\theta^b=0.8\pi$ in  eq.(2.20).
 But after RG transformation, the shape of the potential changes. In 
the first several steps
 ( Region I), the shape still reflects the original form, but after 
$t_{0}$ steps, the shape approaches
 sine curve, which is expected because imaginary coupling of 
 higher IR( irreducible representation) 
dumps much faster
 than the lowest IR one, e.g., ${\rm Im}  \g_{40} \sim ({\rm Im}  \g_{20})^2
\ll |{\rm Im}  \g_{20}|\ll 1$ for any $\theta^b(\not= 0 \quad {\rm 
or}\quad \pi)$. \par
On the other hand, when $\theta^b =\pi$, the shape of $v(x)$ does not change( as is seen 
in eqs.(3. 17)$\sim$ (3.20)).\par  

\Fig( 9) 
\vskip 1cm

\hbox {\bf $\S 4.$ Conclusions and discussions}
\smallskip
Migdal Renormalization Group method is applied to U(2) gauge theory with 
$\theta$-term in 
two space-time
 dimensions. This system is described by two group variables $x$ and $y$,
 corresponding 
to U(1) part and diagonal SU(2)  part of U(2) group respectively.\par
For each $\theta$-parameter, various bare real couplings lead to a unique
 trajectory,
 heat kernel. The trajectory differs, however, depending on bare
 $\theta^b$'s.\par
Bare action is chosen in this paper such that
\item{1)} real part of the action is given by Wilson action corresponding to  
the fundamental representation, i.e., the bare coupling is given by 
$\beta^b_{11}$, 
and all the others are set to be zero.\par
\item{2)} imaginary( $\theta$-term) action is given by ``standard 
$\theta$-term action",\par
$$ v(x)=\alpha x$$
which contains only U(1) variable $x$ and 
$$ \gamma^b_{l_1 l_2}(l_2\not= 0)=0.$$ 
\smallskip
Staring from this bare action, imaginary couplings with non trivial SU(2)
 representations are induced as a result of RG transformations, i.e., due to
 the interference between $\beta_{11}$
 and $\gamma_{l_1 0}$'s.\par
The imaginary couplings with nontrivial SU(2) part( $l_2\not= 0$) , however 
go to zero finally 
after many steps of RG transformation, which will be related to the 
fact that the real 
couplings with
 nontrivial SU(2) part finally go to zero infrared fixed point for any 
$\theta$-parameter
 even for $\pi$.\par  
Real couplings are affected by the imaginary ( $\theta$-term) couplings.\par
 \item{(i)} $\beta_{l_1 0}$'s( any $l_1$) go to non zero fixed point
  for $\theta^b =\pi$.
 These couplings even for
$\theta^b \not= \pi$ are first attracted by the fixed point due to
 $\theta=\pi$ in  the first several RG steps.  Spending some RG steps 
around  it, they  finally start to move to the infrared fixed point
 $\beta_{l_1 0}=0$
 when $\theta^b \not= \pi$. 
The string tension $\sigma_{20}$ coming from the Wilson loop in the IR(2, 0) 
tends 
to zero as $\theta^b \rightarrow \pi$. It means the system undergoes 
deconfinement
 phase transition at $ \theta= \pi$.
 \item{(ii)} Even for  $ \theta^b \rightarrow \pi$, $\beta_{l_1 l_2}(l_2\not= 0)$
 goes to strong coupling limit($ \beta_{l_1 l_2} \rightarrow 0$ ). 
  We also saw that 
$\sigma _{11}$ does
 not go to zero but to a finite value. It means IR(1, 1) does not undergo 
deconfinement phase transition 
  due to the influence of SU(2) part.\par
As the future problems,
\item{1.} The RG study of  $CP^{N-1}$ system with $\theta$-term in 2 
space-time 
dimensions will be quite interesting. It should be made clear whether 
$\theta_c$
 depends on bare real coupling $\beta^b$ and whether 
$\theta_c$ will move to smaller values as $\beta^b \rightarrow $ large.\par
\item{2.} Renormalization group study about four dimensional gauge systems
 ($Z_N$ or U(1)) with $\theta$-term 
will be interesting.
 To study the rich phase structure conjectured by Cardy and Rabinovic based on
 the free energy argument should be confirmed by RG method or numerical
 simulations.\par 
\bigskip

\bigbreak\bigskip
\centerline{{\bf Acknowledgements}}\nobreak
The authors are grateful to members of Theory group at Kyushu University
 for discussions. 
This work is supported by Grant-in-Aid (C) No.08640381(Masahiro Imachi) and
 No.07640417  (Hiroshi Yoneyama   ) from the Ministry of
  Education, Science, Sports and Culture.\par
       

\eject
\centerline{{\bf  Figure Captions}}\nobreak

\item{Fig. 1.}RG flow of real coupling constants for $\theta=0$. Bare
 couplings are $\b^b_{11}=1, \cdots ,8$.
Every flow converges a unique trajectory at second step(scale parameter
  is chosen in this paper as $\lambda^2=2$). Heat kernel is also shown.
 Renormalized coupling is close to heat kernel couplings.\par 
\item{Fig. 2.}The same as Fig. 1. Flow in strong coupling region is shown.
 Independent of bare couplings, all the
points lie on a unique trajectory.\par 

\item{Fig. 3.}Flows for nonzero $\theta$'s are shown. Bare couplings are
 $\beta^b_{11}=4 
\ \ {\rm and}\ \  6$, $\theta=0.8\pi \ \ {\rm and }\ \ 1.0\pi$. Trajectory
 depends on 
bare $\theta$ but
 independent of bare $\beta$ for each $\theta$.\par  
\item{Fig. 4.}Flow of couplings of U(1) part.
Thick line, dashed line and thin line   denotes $\theta=0.8\pi , 
 \theta=0.999\pi$ and $1.0\pi$, respectively.  
 At first( region I) they
 are attracted by the fixed point
 $\b_{20}=1/2, \b_{40}=-1/4$, controlled by $\theta=\pi$.  After 
 spending several steps( $t_0$), they 
start to move( region II) to infrared fixed point $\b_{20}=0,\ \  \b_{40}=0$.
 The influence of $\theta=\pi$ fixed point is stronger for
 bare $ \theta$  close to $\pi$.\par 
\item{Fig. 5.}Flow of couplings of nontrivial representation of SU(2).
 Coupling $\g_{11}$, originally
 zero, acquires nonzero values due to RG effect. It, however, goes to
 to zero by the effect of SU(2) confinement 
fixed point.\par  
\item{Fig. 6(a).}String tensions, $\sigma_{20} \ 
$ (bold line) ${\rm and}
\  \sigma_{11}$
(thin line)
, 
are shown for various bare $\theta$.
 $\sigma_{20}$ goes to zero at $\th=\pi$. It shows IR (2, 0) is 
deconfined by the back ground electric field
 coming from $\theta$-term. $\sigma_{11}$ approaches non zero value even 
at $\th=\pi$. It shows
 IR(1, 1) belongs to non trivial SU(2) representation and stays always
 in confinement phase, since 
there is no back ground field contributing to SU(2) part.\par   
\item{Fig. 6(b).}Free energy vs. $\th$. It is proportional to $\th^2$.\par 
\item{Fig. 7.}Flow of imaginary couplings of U(1) part. Bare $\th$ are 
$0.8\pi 
\ {\rm and}\  0.999\pi$. 
At first several RG steps(region I), renormalized couplings  are attracted by
 the non trivial
 fixed point( $\th=\pi$).  Staying at the starting 
point for a number of steps,  
they then begin to move to infrared fixed point.
 The effect of the nontrivial fixed point is stronger for $\th$ near to 
$\pi$.\par  
\item{Fig. 8.}Flow of imaginary couplings, ${\rm Im} \g_{11} \ {\rm and}
\  {\rm Im} \g_{22}$,
 with nontrivial SU(2) part.
In region I, absolute values of them increase but they move to infrared
 fixed point finally.\par 
\item{Fig. 9.}Change of shape of imaginary action under RG transformations.
Here $t$ denotes the number of steps of renormalization group transformation. 
 At first, linear $x$ dependence is clear,
 but afterwards shape changes to sine curve. This is due to the fact that
 only  single  IR( the lowest representation) survives in strong coupling 
regions.\par
  
\listrefs
\input epsf
$$\vbox{ 
\vskip 2cm
\centerline{Figure 1}
\bigskip
\epsfysize=0.7\hsize
\hskip  -1.5 cm
\epsfbox{fig1.epsf}
}$$
\smallskip
\vfill
\eject
$$\vbox{ 
\vskip 3cm
\centerline{Figure 2}
\bigskip
\hskip -2.5 cm
\epsfysize=.7\hsize
\epsfbox{fig2.epsf}
}$$
\smallskip
\vfill
\eject
$$\vbox{ 
\vskip 2cm
\centerline{Figure 3}
\bigskip
\epsfysize=0.7\hsize
\hskip 0.0cm
\epsfbox{fig3.epsf}
}$$
\smallskip
\vfill
\eject
$$\vbox{ 
\vskip 2cm
\centerline{Figure 4}
\vskip 2cm
\epsfysize=.5\hsize
\hskip 0cm
\epsfbox{fig4.epsf}
}$$
\smallskip
\vfill
\eject
$$\vbox{ 
\vskip 2cm
\centerline{Figure 5}
\bigskip
\epsfysize=.7\hsize
\hskip-0.5cm
\epsfbox{fig5.epsf}
}$$
\smallskip
\vfill
\eject
$$\vbox{ 
\vskip 1.0cm
\centerline{Figure 6(a)}
\bigskip
\hskip.2cm
\epsfysize=.5\hsize
\epsfbox{fig6a.epsf}
\vskip 1cm
\centerline{Fig. 6(b)}
\bigskip
\hskip1.0cm
\epsfysize=.6\hsize
\epsfbox{fig6b.epsf}
}$$
\smallskip
\vfill
\eject
$$\vbox{ 
\vskip 2cm
\centerline{Figure 7}
\bigskip
\epsfysize=.6\hsize
\hskip.05cm
\epsfbox{fig7.epsf}
}$$
\smallskip
\vfill
\eject
$$\vbox{ 
\vskip 2cm
\centerline{Figure 8}
\vskip 2cm
\epsfysize=.5\hsize
\hskip -0.5cm
\epsfbox{fig8.epsf}
}$$
\smallskip
\vfill
\eject
$$\vbox{ 
\vskip 2cm
\centerline{Figure 9}
\bigskip
\epsfysize=.9\hsize
\hskip.05cm
\epsfbox{fig9.epsf}
}$$
\smallskip
\vfill
\eject
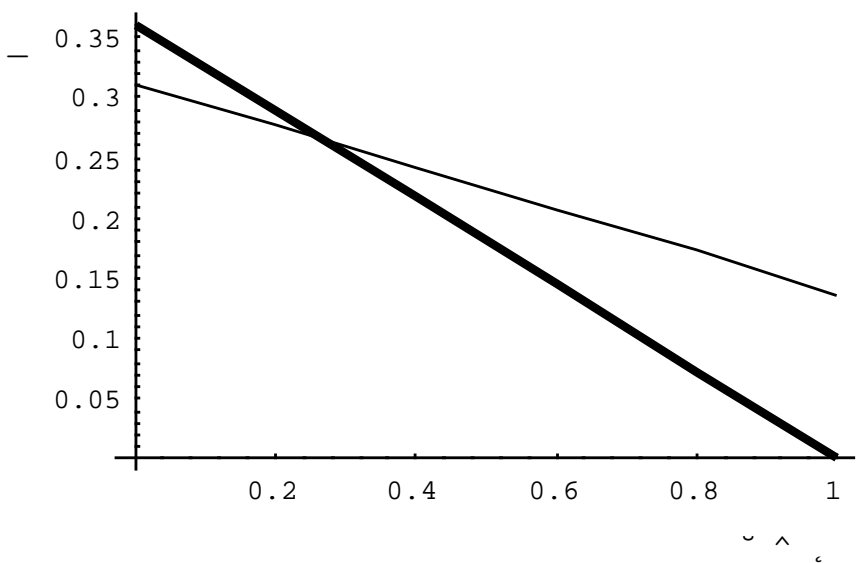
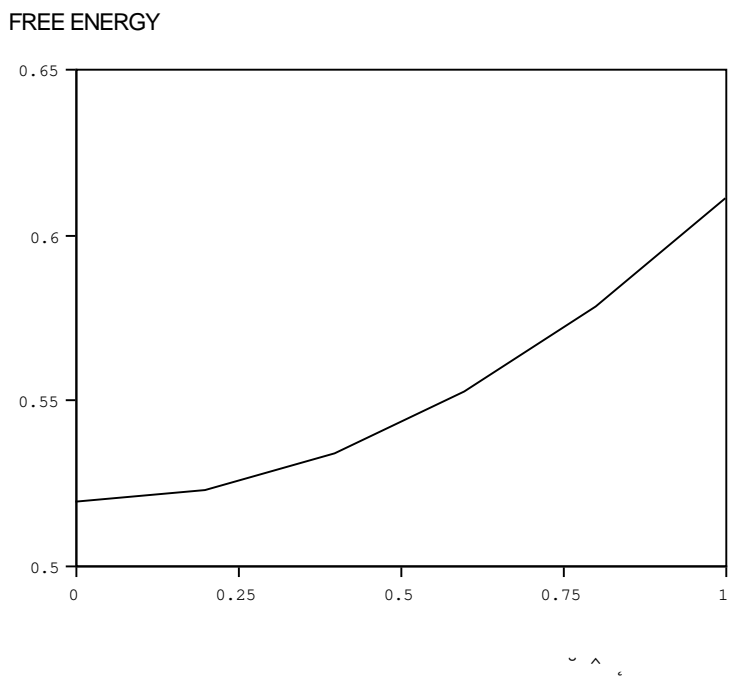
\bye